# Superior enhancement in thermal conductivity of epoxy/graphene nanocomposites through use of dimethylformamide (DMF) relative to acetone as solvent

Swapneel Danayat, Avinash Singh Nayal, Fatema Tarannum, Roshan Annam, Rajmohan Muthaiah and Jivtesh Garg

School of Aerospace and Mechanical Engineering, University of Oklahoma, Norman, OK 73019, USA.

Email ID: sdanayat@ou.edu, Avinash.Singh.Nayal-1@ou.edu, fatema-1@ou.edu, anna0003@ou.edu, rajumenr@ou.edu, garg@ou.edu,

**Abstract:** In this work**,** we demonstrate that use of dimethylformamide (DMF) as a solvent leads to better dispersion of graphene nanoplatelets in epoxy matrix compared to acetone solvent, in turn leading to higher thermal conductivity epoxy-graphene nanocomposites. While role of solvents in enabling superior mechanical properties has been addressed before, outlined study is the first to address the effect of solvents on thermal conductivity enhancement and provides novel pathways for achieving high thermal conductivity polymer composite materials. Uniform dispersion of graphene nanoparticles into epoxy can improve thermal contact with polymer leading to superior interface thermal conductance between polymer matrix and graphene. Organic solvents are typically employed to achieve efficient dispersion of graphene into the epoxy matrix. In this study, we compare the effect of two organic solvents, dimethylformamide (DMF) and acetone, in terms of their efficiency in dispersing graphene into the epoxy matrix and their effect on enhancing thermal conductivity of the composite. We find that polymer-graphene composites made with DMF solvent show 44% higher thermal conductivity compared to those made using acetone at 7 weight% filler composition. Laser scanning confocal microscopy (LSCM) imaging reveals that graphene-epoxy composites, prepared using DMF as solvent, exhibit more uniform dispersion of graphene-nanoplatelets compared to the case of acetone with acetone-based samples exhibiting up to 211% larger graphene agglomerations. Comparison with effective medium theory reveals an almost 35% lower interface thermal resistance between graphene and epoxy for DMF relative to acetone prepared composite. These results provide fundamentally new avenues to achieve higher thermal conductivity graphene-epoxy composites, of key importance for a wide range of thermal management technologies.

**1. Introduction:** Thermal management has become a challenging issue in modern electronics due to continuous miniaturization of electronic components which results in increasing heat fluxes. To improve the efficiency and reliability of electronic systems, heat needs to be dissipated efficiently(1, 2). In terms of material selection, polymers offer several advantages over metals such as low cost, corrosion resistance, easy of moldability, and lower weight. High thermal conductivity polymer materials can improve thermal management in a wide range of applications, such as - water desalination(3), automotive control units(4), batteries(5), solar panels(6), supercapacitors(7), electronic packaging(8), and electronic cooling(9). A key approach to enhance thermal conductivity of polymers is addition of high thermal conductivity fillers such as graphene ($k$ >2000 Wm$^{-1}$K$^{-1}$ (10, 11)). Different approaches have been used to enhance composite $k$ value through graphene, such as synergistic effect with multiple fillers(12, 13) and alignment of graphene(14-16). The success of these approaches is, however, limited by the large interface thermal resistance between graphene and polymer in the range of $10^{-8}$ to $10^{-7}$ m$^2$ KW$^{-1}$ (16, 17) due to mismatch of phonons (lattice vibrations) between the two. Agglomeration of graphene can increase effective interfacial thermal resistance by preventing wetting of graphene particles with polymer. Interfacial thermal conductance can be improved through more uniform dispersion of graphene nanoplatelets into epoxy matrix. Typically, in preparation of epoxy/graphene nanocomposites, solvents such as acetone and DMF are used to aid dispersion of graphene in to the epoxy matrix. These solvents can have large differences in their ability to promote uniform dispersion of graphene, potentially leading to significant differences in resulting thermal conductivity enhancement. In the past, epoxy/graphene nanocomposites used for thermal conductivity research have always been prepared using acetone as the solvent(18-20). In this work, we demonstrate for the first time that use of DMF as solvent can lead to 44% higher thermal conductivity compared to the use of acetone at 7% filler composition. Outlined results provide fundamentally novel pathways for achieving high thermal conductivity epoxy/graphene nanocomposites through use of more effective solvents for dispersion of graphene.

Graphene, since its discovery in 2004(21), has been thoroughly studied due to its extraordinary properties like extremely high electrical resistivity, high thermal conductivity, high mechanical strength, and modulus (22). This study focuses on the role of solvents in enabling superior dispersion of graphene in epoxy composites for achieving higher thermal conductivity values. High thermal conductivity of graphene (2000 Wm$^{-1}$K$^{-1}$ – 5000 Wm$^{-1}$K$^{-1}$) makes it an ideal filler material for obtaining thermally conducting polymer composites (23-27). Incorporation of graphene nanoplatelets into polymer matrix has been shown to yield significant enhancement in physical properties of polymers (28). Uniform dispersion of graphene into the polymer is of utmost importance to achieve maximum improvement in thermal conductivity. Uniform dispersion leads to superior thermal contact between graphene nanoplatelets and polymer matrix leading to lower thermal interface resistance between the two, thus enabling higher composite thermal conductivity. Uniform dispersion can also further enhance thermal conductivity by allowing formation of efficient graphene percolation networks in the polymer matrix through a reducing in gap between the graphene nanoparticles (28-30).

Commonly used polymers in engineering applications are mostly soluble in organic solvents. This makes dispersion of graphene sheets into organic-solvents, an important parameter to prepare a homogeneous composite. Graphene also cannot be simply mixed mechanically with the polymer as it tends to reaggregate due to strong Van der Waals forces between graphene sheets. Furthermore, high viscosity of typical polymer melts also prevents direct uniform dispersion of graphene into polymer. Therefore, using organic solvents compatible with the common polymers is typically employed for uniformly dispersing graphene sheets(29-31).

Studies have been conducted for the effect of solvents on dispersion of graphene in regards with improvement of mechanical properties (32). However, no results have been reported for the effect of solvents on improvement of thermal conductivity. Acetone, one of the commonly used solvent for preparing graphene-polymer composites shows shorter-term stability of graphene dispersions when compared to many other organic solvents like dimethylformamide (DMF), tetrahydrofuran (THF), N-methyl-2-pyrrolidone (NMP) and ethylene glycol, which show long term stability of dispersions (33). The improved dispersion of graphene in solvents such as DMF compared to acetone, has been shown to also lead to improved mechanical properties such as mechanical strength. Mishra *et al.* (34)compared the effect of three solvents (ethanol, acetone, and toluene) on the dispersion of polyhedral oligomeric silsequioxane (POSS) in epoxy and the subsequent enhancement of mechanical properties of POSS-epoxy nanocomposite. They found that composite prepared using ethanol showed increase in elastic modulus and fracture toughness values due to the better dispersion of POSS in epoxy resin in ethanol solvent. While above studies addressed the effect of improved dispersion in certain solvents on mechanical properties like mechanical strength, the present study focuses on the effect of solvents on thermal conductivity.

Previous works have reported thermal conductivity enhancement of epoxy/graphene nanocomposites prepared using acetone as solvent. Wang et al prepared graphene nanoplatelet/epoxy samples using acetone and achieved 0.45 $Wm^{-1}K^{-1}$ thermal conductivity at 5wt% filler loading(18). Guo et al demonstrated a thermal conductivity close to 0.3 $Wm^{-1}K^{-1}$ and 0.45 $Wm^{-1}K^{-1}$ for graphene/epoxy composite samples made with acetone at 5wt% and 10wt% loading respectively(19). S Han et al showed thermal conductivity of 0.33 $Wm^{-1}K^{-1}$ for epoxy/graphene nanoplatelets at 4wt% loading prepared using acetone(20). Multiple such studies reporting thermal conductivity enhancement of graphene/epoxy composites prepared using acetone exist in literature. There are, however, no reported studies on thermal conductivity of epoxy/graphene composites prepared using DMF as solvent and this work aims to fill that gap and provide more efficient pathways to achieve high thermal conductivity graphene/epoxy composites.

The dispersion capability of a solvent can be inferred from their 'Hansen Solubility Parameters' (HSP). These parameters include a dispersion cohesion parameter ($\delta_d$), polarity cohesion parameter ($\delta_p$) and a hydrogen bonding cohesion parameter ($\delta_h$) (35). S. Park et al showed that highly reduced graphene (HRG) was well dispersed in solvent mixtures (DMF/$H_2O$ mixed with either acetone, acetonitrile, THF, DMF, NMP, DMSO and ethanol) having the sum $\delta_p+\delta_h$ in the range of 13~29. They further showed that the solvent having the sum lower than 10 (DMF/$H_2O$ mixed with either DCB, diethyl ether and toluene) and higher than 30 (DMF/$H_2O$ mixed with water) exhibit poor dispersion of HRG. The HSP sum ($\delta_p+\delta_h$) of acetone and DMF are 17.4 and 25 respectively, implying they both are viable alternatives for dispersing HRG(36).

Another important parameter related to dispersion is the zeta potential (ς). Zeta potential is the electric potential difference between the dispersion medium and the stationary layer of fluid attached to the dispersed particle(37). It is denoted using a numerical value with a positive or negative sign. A strong positive or negative zeta potential result in high repulsive forces between the dispersed particle, indicating good stability of dispersions(38). A. A. S. Ghazvini et al have shown that the zeta potential for graphene in acetone is -22.6 mV(39). S. Gambhir et al have shown that graphene in different forms has zeta potential lower than -30mV in DMF(40). These values suggest a stronger negative zeta potential of graphene in DMF which leads to better stability of the graphene dispersion in DMF over time. Villar- Rodil et al have demonstrated stable suspension of unreduced and chemically reduced graphene in DMF over several months(31). We also performed stability tests for dispersion of graphene in DMF and acetone by separately suspending graphene in DMF and acetone and observing the suspension as a function of time. These results are reported in section 4.2.

The present study is conducted for three graphene concentrations- 3 weight%, 5 weight% and 7 weight%. The dispersion effect is characterized using laser scanning confocal microscopy which provides optical images of the graphene sheets dispersed in the polymer matrix, enabling a visual comparison of achieved dispersion. We first present materials preparation followed by thermal conductivity results and discussion.

**2.Materials:** Graphene nanoplatelets were purchased from Graphene Supermarket. Characterization of graphene is provided in the supplementary information. Epoxy resin used for the study was EPIKOTE RESIN MGS RIMR 135 and the hardener used was EPIKURE CURING AGENT MGS RIMH 137, both purchased from Hexion. The organic solvents N-N, Dimethylformamide (DMF) and Acetone were purchased from University of Oklahoma chemical stock room.

**3. Experimental Work**

*3.1 Preparation of Epoxy/Graphene composite with Acetone:* Graphene nanoplatelets (3-7 nm thick and ~5 microns in lateral size) were dispersed in 80 mL acetone and tip sonicated for one hour in an ice bath to prevent heating and evaporation of acetone. The resin was then added to the solution and tip sonicated in an ice bath for another two hours. After sonication, the solution was heated to 80 °C and stirred continuously using a mechanical mixer to remove the solvent. The mixture of graphene and epoxy was weighed until required weight was reached related to removal of acetone. This mixture was then spread on a PTFE sheet and placed in a vacuum oven at 65 °C for 15 hours to further ensure complete removal of acetone. The composite mixture was then transferred into aluminum molds and cured at 90 °C for 20 hours.

*3.2 Preparation of Epoxy/Graphene composite with DMF:* Graphene nanoplatelets (3-7 nm thick and ~5 microns in lateral size) were dispersed in 80 mL DMF and bath sonicated in an ice bath for 30 minutes. Epoxy resin was then added to the solution and bath sonicated for another one hour.

After sonication, the solution was heated and stirred at 150 °C to remove the solvent. The graphene-epoxy mixture was then spread on a PTFE sheet and placed in a vacuum oven at 140 °C for 15 hours to ensure complete removal of the solvent. The mixture was then transferred into aluminum molds and cured at 90 °C for 20 hours. Optical image of the prepared composite samples is shown in figure 1.

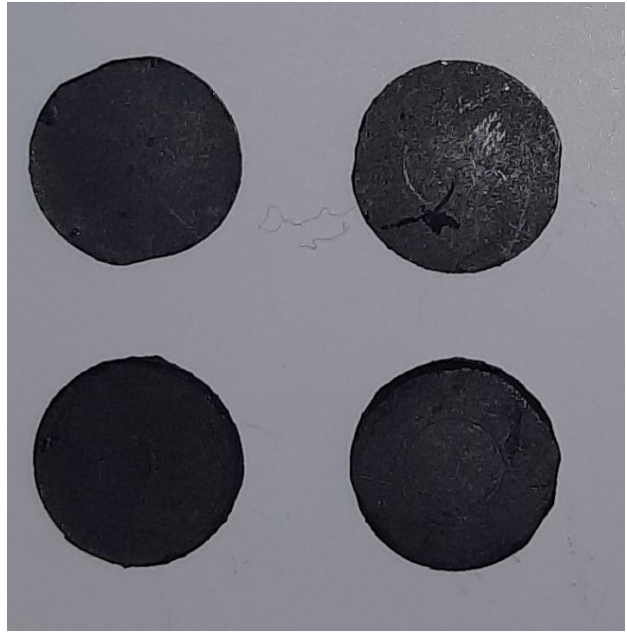

Figure 1. Optical images of epoxy/graphene composite samples.

*3.3 Measurements and Characterizations:* Thermal conductivity of the composite samples was calculated using the thermal diffusivity values measured using the NETZSCH LFA 467 which works on the principle of laser flash analysis. In this process, a light pulse beam heats the lower surface of the sample and the temperature increase on the upper surface of sample is measured using an infrared detector. The temperature rise on the upper surface of sample is recorded as a function of time. Thermal diffusivity is then calculated using the equation

$$\alpha = 0.1388 \frac{d^2}{t_{\frac{1}{2}}} \qquad (1)$$

where $\alpha$ is the thermal diffusivity, $d$ is the sample thickness and $t_{\frac{1}{2}}$ is the time taken to reach the half of the maximum temperature. Using this thermal diffusivity, the thermal conductivity is calculated using the equation

$$k = \alpha \rho C_p \qquad (2)$$

where $k$ is the thermal conductivity, $\rho$ is the density and $C_p$ is the specific heat of the sample respectively.

## 4. Results and Discussion:

### *4.1 Thermal Conductivity Measurements:*

Figure 2 shows the thermal conductivity values of the graphene-epoxy composites with 3wt% (1.54 vol%), 5wt% (2.6 vol%) and 7wt% (3.68 vol%) concentration for the two organic solvents - acetone and DMF. Thermal conductivity of pure epoxy sample is measured to be 0.17 Wm$^{-1}$K$^{-1}$. At 3wt%, DMF and acetone samples show identical thermal conductivity value of 0.34 Wm$^{-1}$K$^{-1}$. However, at higher concentrations, DMF-based composite samples show significantly higher thermal conductivity values relative to acetone-based composite. At 5wt% and 7wt% graphene concentrations, DMF based composite exhibit $k$ values of 0.63 Wm$^{-1}$K$^{-1}$ and 0.9 Wm$^{-1}$K$^{-1}$ respectively, while acetone-based composite demonstrate lower $k$ values of 0.42 Wm$^{-1}$K$^{-1}$ and 0.55 Wm$^{-1}$K$^{-1}$ respectively. The $k$ values of DMF based composite are approximately 40% and 44% higher relative to those of acetone-based composite at 5wt% and 7wt% graphene concentrations respectively. These results for the first time highlight the distinct advantage of DMF as a solvent in achieving higher thermal conductivity epoxy/graphene nanocomposites.

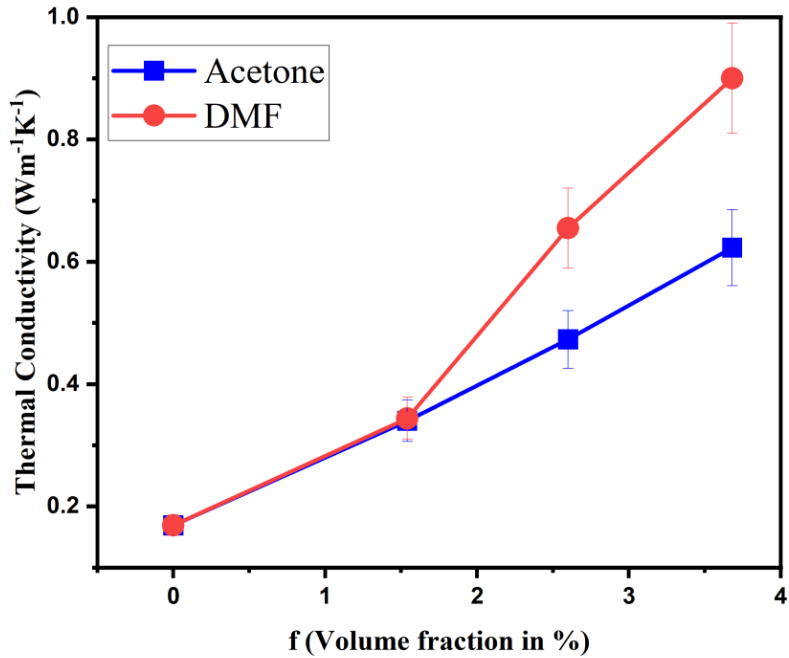

Figure 2. Thermal conductivity variation with increasing graphene concentration

We next demonstrate, using laser scanning confocal microscopy (LSCM), that this beneficial impact of DMF is due to the more uniform dispersion of graphene in the resulting graphene/epoxy nanocomposite.

*Laser Scanning Confocal Microscopy (LSCM) Imaging for Dispersion Characterization :*

The dispersion effect of the two solvents was characterized using laser scanning confocal microscopy (LSCM) which is an optical imaging technique. LSCM is powerful imaging technique that uses a spatial pinhole to block out the out-of-focus light thus increasing the optical resolution of a micrograph. LSCM generate images with lesser haze and better contrast than a conventional microscope and can be focused on a thin cross-section of a sample (14, 16, 41). LSCM generates optical sections as thin as 300 nm axially by using reflected light to create images. By collecting a series of these optical sections along the optical axis, one can generate a 3D reconstruction of a volume within an intact specimen. This makes LSCM an effective tool to visualize the dispersion of graphene within the composite samples. In this work we used a Leica SP8 laser scanning confocal microscope with a 561 nm diode pumped solid state (DPSS) laser. The samples are imaged with a 63 × 1.4 oil immersion objective with a pinhole aperture at 0.2 airy units (AU) and voxel dimensions of 120 nm × 120 nm × 120 nm.

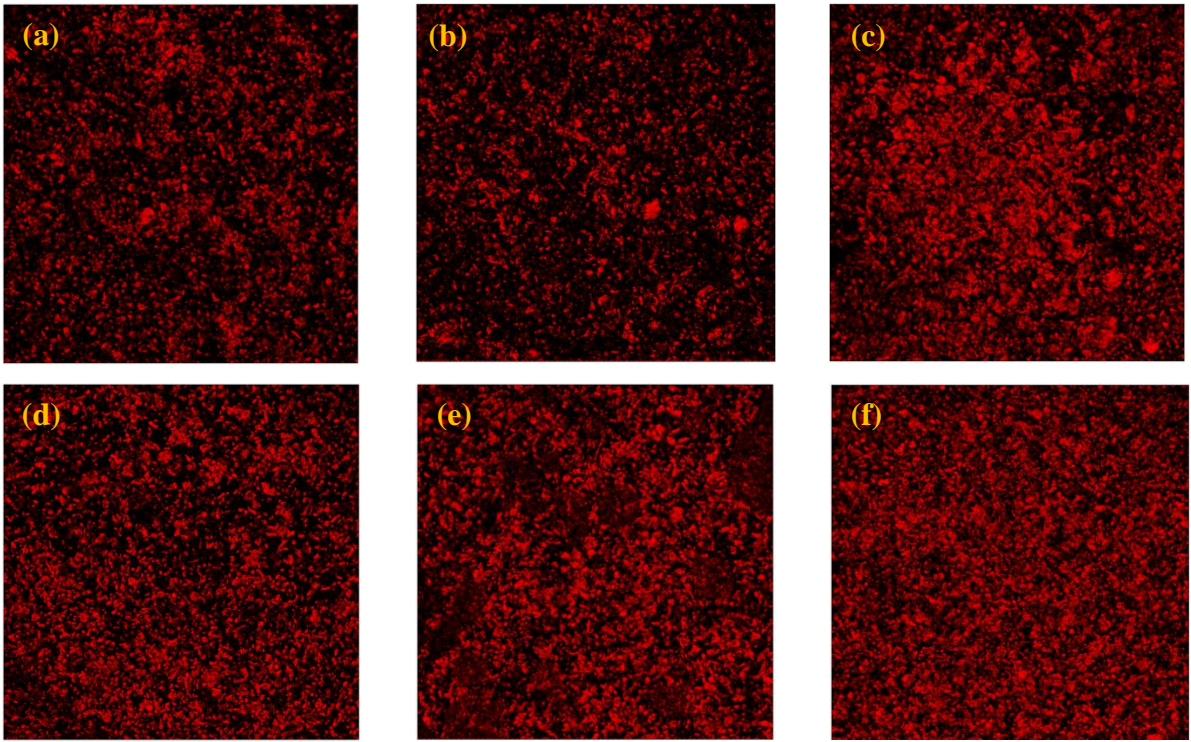

Figure 3. Confocal microscopy images for acetone-based samples for concentrations a) 3wt%, b) 5wt%, c) 7wt% and DMF-based samples for concentrations d) 3wt%, e) 5wt%, f) 7wt%

Figure 3 shows the LSCM images of the graphene-epoxy composites with 3 wt%, 5 wt% and 7 wt%. The red color represents graphene nanoplatelets dispersed within the composite. Figs. 3a, b and c show graphene distribution within the acetone based samples, while Figs. 3d, e and f are for the DMF based samples for the 3, 5 and 7 wt% composition respectively. There is clear visible difference in the dispersion of graphene in the epoxy composites prepared using DMF and acetone solvents. Acetone based samples (Fig. 3 a,b,c) show larger gaps between graphene nanoparticles indicating poor dispersion. Composites prepared using DMF, however, show relatively more uniform distribution of graphene sheets (Fig. 3 d,e,f).

LSCM further allows quantitative analysis of size of particles embedded within the matrix. Figure 4 shows the maximum graphene agglomeration size comparison between the acetone and DMF based samples. The calculation of agglomeration size was performed using the Fiji (also known as

ImageJ) image analysis software. The confocal microscopy images were analyzed for 3D particle volumes using the software. Results show that at 3wt% composition, the agglomerate sizes are identical which explains the similar thermal conductivity values at this concentration. At higher concentrations, agglomerate size in acetone-based samples are 211% and 93% higher at 5wt% and 7wt% respectively compared to DMF based samples. The significantly lower agglomerate sizes in DMF-based samples provide clear evidence of the more uniform dispersion of graphene nanoplatelets in the DMF based composites.

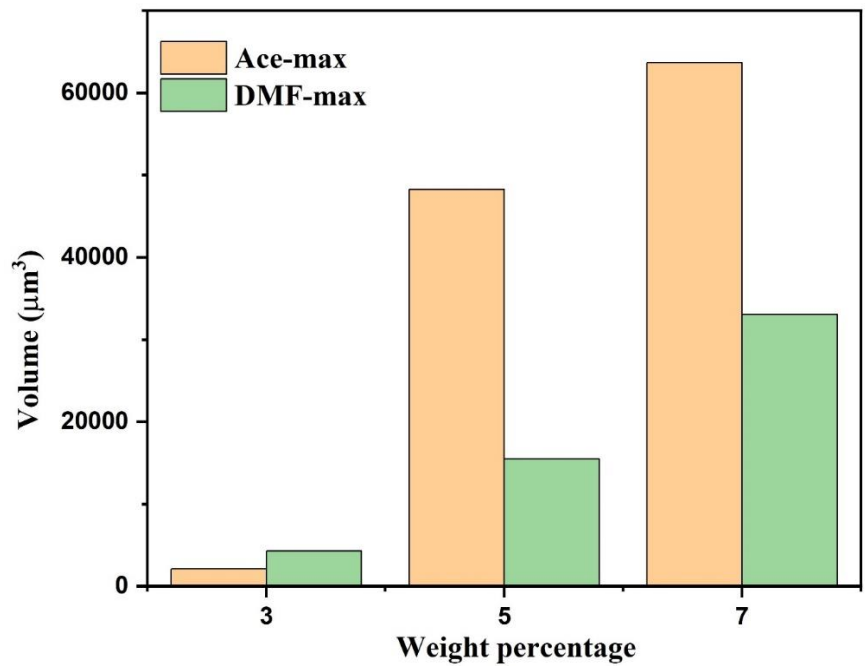

Figure 4. Maximum graphene agglomeration volume comparison

To understand the stability of graphene nanoplatelets in the two solvents, we prepared separate dispersions of graphene nanoplatelets in acetone and DMF and recorded the state of dispersion after regular intervals of time. These recorded images at time intervals of 0, 5 and 24 hours are shown in Fig. 5. The more stable dispersion of graphene in DMF is clearly visible in Fig. 5 which shows that while in acetone, graphene nanoplatelets completely sediment after 24 hours, they are still suspended in DMF after the same time interval.

LSCM analysis combined with stability tests provided clear evidence of the role of DMF solvent in facilitating a more uniform dispersion of graphene in the epoxy matrix.

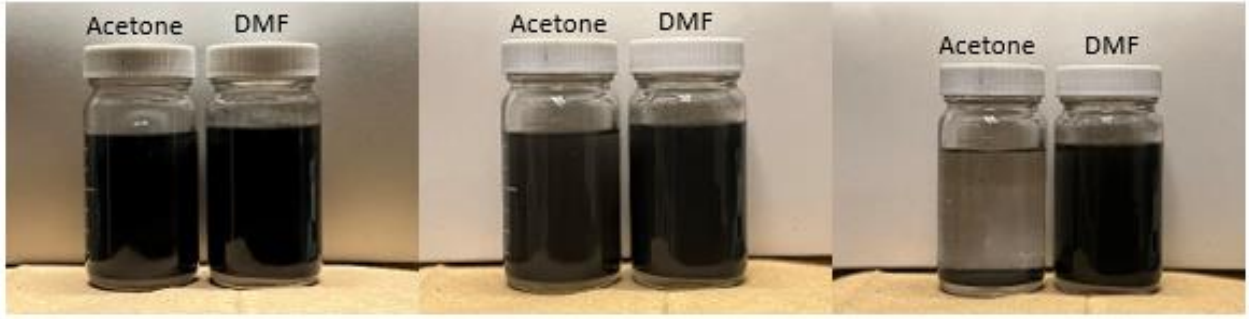

Figure 5. Graphene suspension in acetone and DMF at t=0,5 and 24 hours

*4.3 Effective Medium Theory*: The effect of graphene dispersion on the thermal conductivity of graphene-epoxy nanocomposite was theoretically studied using the effective medium theory presented by Nan et al (14, 42). The theory takes into account the effect of interfacial thermal resistance between the dispersed particles and the matrix material on the overall composite thermal conductivity. Interfacial thermal resistance is an essential parameter in determining the thermal conductivity of any composite material. Using the theory presented by Nan et al and the measured thermal conductivity values, we calculate the interface thermal resistance for acetone and DMF based samples.

The theoretical effective thermal conductivity of the graphene-epoxy composites is given by

$$k_{effective} = k_m \frac{2 + f[\beta_{11}(1 - L_{11})(1 + \langle \cos^2 \theta \rangle) + \beta_{33}(1 - L_{33})(1 - \langle \cos^2 \theta \rangle)]}{2 + f[\beta_{11}L_{11}(1 + \langle \cos^2 \theta \rangle) + \beta_{33}L_{33}(1 - \langle \cos^2 \theta \rangle)]} \quad (3)$$

where $k_{effective}$ & $k_m$ is the effective thermal conductivity of composite and thermal conductivity of pristine epoxy matrix respectively for $f$ volume fraction of graphene nanoparticles. The $\langle \cos^2 \theta \rangle$ term considers the orientation of the filler material. For the present scenario, a random orientation of graphene nanoparticles is considered ($\langle \cos^2 \theta \rangle$=1/3).

The geometrical parameters of the oblate graphene nanoparticles such as aspect ratio $p$ are considered by the equations

$$L_{11} = L_{22} = \frac{p^2}{2(p^2 - 1)} + \frac{p}{2(1 - p^2)^{3/2}} \cos^{-1} p \quad (4)$$

$$L_{33} = 1 - 2L_{11}$$

In equation (3), $\beta_{ii}$ is computed using the equation

$$\beta_{ii} = \frac{K_{ii}^c - k_m}{k_m + L_{ii}(K_{ii}^c - k_m)} \quad (5)$$

where $K_{ii}^c$ are the effective thermal conductivity values of the graphene nanoparticles considering the effect of thermal interface resistance. The in-plane effective values $K_{11}^c$, $K_{22}^c$ and the through-plane effective value $K_{33}^c$ are given by equations

$$K_{11}^c = K_{22}^c = \frac{k_{p1}}{1 + \gamma L_{11}\, k_{p1}/k_m} \qquad (6)$$

$$K_{33}^c = \frac{k_{p3}}{1 + \gamma L_{33}\, k_{p3}/k_m} \qquad (7)$$

$$\gamma = (1 + 2p)\alpha \qquad (8)$$

$$\alpha = \frac{R k_m}{t} \qquad (9)$$

where $R$ is the thermal interface resistance and $k_{p1}$ and $k_{p3}$ are the in-plane and through-plane thermal conductivity values of graphene respectively.

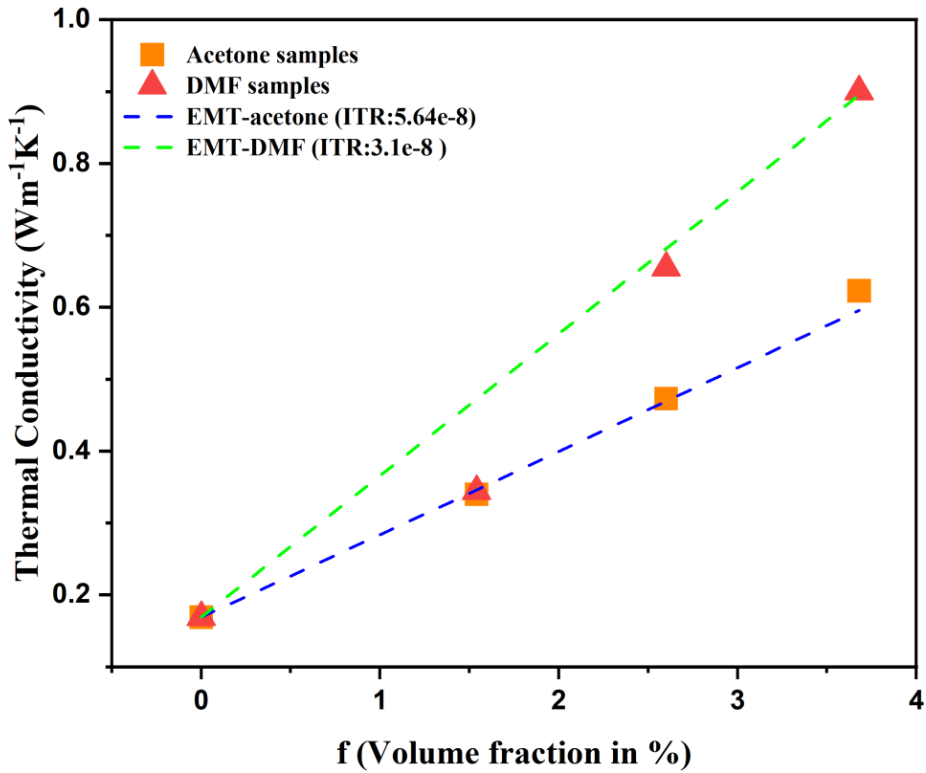

Figure 6. Effective medium theory and measured thermal conductivity values (ITR: Interfacial Thermal Resistance)

Figure 6 shows the predicted thermal conductivity values from the effective medium theory in comparison to measured values. Respective interface thermal resistance resistances for the acetone and DMF based samples were calculated by forcing good agreement between measured and predicted thermal conductivity values. Results show that the interface thermal resistance (ITR) for acetone-based samples is 82% higher than the DMF based samples. This difference in the thermal interface resistance is key in understanding the higher thermal conductivity of the DMF based samples.

## 5. Conclusion :

In summary, this study compares the effect of two solvents, namely, DMF (dimethylformamide) and acetone on dispersion of graphene into epoxy matrix and subsequent enhancement of thermal conductivity of the graphene-epoxy nanocomposite. Epoxy/graphene nanocomposites were prepared using two solvents – DMF and acetone to facilitate dispersion of graphene into the epoxy matrix. Measurements revealed that DMF based composites exhibited 40% and 44% higher thermal conductivity than the acetone-based samples at 5wt% and 7wt% graphene content respectively. This significantly higher thermal conductivity in DMF based samples, was found to be due to more uniform dispersion of graphene nanoplatelets in the epoxy matrix for the DMF based samples. Laser Scanning Confocal Microscopy was used to obtain optical images of the composite samples to visualize the dispersion of graphene in the epoxy matrix. Analysis revealed almost 211% and 93% smaller graphene agglomerate size in DMF based samples relative to acetone-based samples, clearly indicating better dispersion of graphene in DMF prepared composites. Finally effective medium theory was used to extract the interface thermal resistance between graphene and epoxy, for the two samples. Interface thermal resistance for DMF based samples was found to be almost 82% lower relative to acetone-based samples due to the improved dispersion of graphene in samples prepared using DMF solvent. These results provide fundamentally novel pathways for developing next generation high thermal conductivity graphene-epoxy nanocomposites through solvent induced improved dispersion of graphene.

**Acknowledgements:** We acknowledge Dr. Tingting Gu (University of Oklahoma) for providing access to and assisting with the Laser Scanning Confocal Microscopy.

**Funding:** We acknowledge financial support from National Science Foundation CAREER award under Award No. 1847129.

**Declaration of interests:** The authors declare that they have no known competing financial interests or personal relationships that could have appeared to influence the work reported in this paper.